\documentclass[%
 aip,
rsi,%
 amsmath,amssymb,
preprint,%
]{revtex4-1}

\usepackage{graphicx}
\usepackage{bm}

\usepackage{color}

\begin{document}


\newcommand{\be}{\begin{equation}} 
\newcommand{\ee}{\end{equation}}
\newcommand{\bea}{\begin{eqnarray}}   
\newcommand{\eea}{\end{eqnarray}}

\newcommand{\rr}{{\bf r}}

\newcommand{\nvec}{\boldsymbol{n}}
\newcommand{\txi}{\boldsymbol{\xi}^t}
\newcommand{\F}{\boldsymbol{F}}
\newcommand{\colxi}{\boldsymbol{u}}

\newcommand{\UU}{{\cal U}}
\newcommand{\xchi}{\boldsymbol{\chi}}
\newcommand{\pvec}{\boldsymbol{p}}
\newcommand{\eeta}{\boldsymbol{\eta}}
\newcommand{\xxi}{\boldsymbol{\xi}}

\newcommand{\xx}{\boldsymbol{x}}
\newcommand{\vv}{\boldsymbol{v}}
\newcommand{\uu}{\boldsymbol{u}}
\newcommand{\xib}{\boldsymbol{\xi}}

\date{\today}
\title{Active escape dynamics: the effect of persistence on barrier crossing }
\author{Lorenzo Caprini}
\affiliation{Gran Sasso Science Institute (GSSI), Via. F. Crispi 7, 67100 L'Aquila, Italy}

\author{Umberto Marini Bettolo Marconi}
\affiliation{ Scuola di Scienze e Tecnologie, 
Universit\`a di Camerino, Via Madonna delle Carceri, 62032, Camerino, INFN Perugia, Italy}

\author{Andrea Puglisi}
\affiliation{CNR-ISC, Consiglio Nazionale delle Ricerche, 
Dipartimento di Fisica, Universit\`a La Sapienza, 
P.le A. Moro 2, 00185 Rome, Italy}

\author{Angelo Vulpiani}
\affiliation{Dipartimento di Fisica, Universit\`a di Roma Sapienza, I-00185, Rome, Italy}

\begin{abstract}
We study a system of non-interacting active particles, propelled by
colored noises, characterized by an activity time $\tau$, and confined
by a double-well potential. A straightforward application of this
system is the problem of barrier crossing of active particles, which
has been studied only in the limit of small activity. When $\tau$ is
sufficiently large, equilibrium-like approximations break down in the
barrier crossing region. In the model under investigation, 
it emerges a sort of ``negative temperature'' region, and numerical
simulations confirm the presence of non-convex local velocity
distributions. We propose, in the limit of large $\tau$, approximate
equations for the typical trajectories which
successfully predict many aspects of the numerical results. The local
breakdown of detailed balance and its relation with a recent
definition of non-equilibrium heat exchange is also discussed.
\end{abstract}

\date{\today}	
\pacs{Valid PACS appear here}

\maketitle



\section{Introduction}

 Recently, there has been an upsurge of interest towards active matter,
namely systems of particles able to convert energy
from the environment into directed persistent motion.  Examples range
from bacterial colonies, spermatozoa to Janus self-propelled particles
\cite{ramaswamy2010mechanics,bechinger2016active,marchetti2013hydrodynamics,romanczuk2012active}.
The propulsion mechanism is realized in different ways: living systems exploit 
metabolic processes in order to move, while artificial particles immersed in a solvent exploit
a chemical reaction catalyzed on their surface.  Among the various models
introduced to describe the behavior of active systems the so-called
active Ornstein-Uhlenbeck particle (AOUP) model \cite{szamel2014self}
occupies an important place because it allows with a minimal set of
ingredients to reproduce some characteristic features of
self-propelling systems and provides a direct and useful bridge
towards the world of colloidal particles.  In practice, the AOUP is
designed to account for the persistence of the trajectories by means
of a random Gaussian forcing term, which is identified with the active
force.  Such a forcing is assumed to have a finite correlation time
$\tau>0$, identified with the persistence time, and a finite amplitude
which is a measure of the degree of activity of the system.  In spite
of the fact that a more realistic modeling of active systems requires
the description of the self-propulsion in terms of non-Gaussian active
stochastic processes such as the active Brownian particle (ABP) model \cite{ebeling1999active,cates2013active}, a great deal of explicit analytical
results has been possible by the use of AOUP~\cite{fodor2018statistical}. The model is able to
reproduce interesting phenomena such as the accumulation of active
particles near purely repulsive boundaries and the motility induced
phase separation (MIPS) \cite{cates2015motility,maggi2015multidimensional,fodor2016far,fily2017equilibrium,caprini2018active}.

Interestingly, there exist a series of results concerning the steady
state behavior of the AOUP model which have been obtained by applying
an adiabatic approximation (the so-called unified colored noise
approximation (UCNA)
\cite{hanggi1995colored,maggi2015multidimensional}) to the governing
equation for the probability distribution function.  The important
outcome of this approximation is the possibility of writing explicitly
the stationary prbability distribution function (PDF) in principle for
any type of potential of convex type, i.e. for all potentials whose
Hessian is positive definite.  The UCNA explains the accumulation and
aggregation phenomena in terms of a decreased potential-dependent
effective mobility of the particles.  If the convexity condition is
not fulfilled one can still use the UCNA for sufficiently small values
of $\tau$, but when the effective mobility becomes negative the
approximation ceases to be valid. On the other hand, one may ask the
following question: how does an active system subject to colored noise
behaves in the presence of a non-convex potential? The question might
seem academic, but on the contrary, this is a situation that certainly
occurs in practice: apart from the cases of some power law confining
potentials or inverse power law purely repulsive potentials, the
forces experienced by active particles might be associated with
non-convex potentials, corresponding to attractive interactions.
Another paradigmatic example is an active particle crossing a barrier,
i.e. in the presence of a bistable potential. This example has been
studied in the recent literature but only in the limit of small
activity \cite{sharma2017escape}.

In the next Section we present the AOUP model in the case of a single particle and recall the main results concerning the steady distribution function which have been obtained in the framework of the UCNA. In Section \ref{section3}  we illustrate the phenomenology
of the AOUP in the presence of a bistable potential when the amplitude of the active force and the persistence time are large.
One observes that basically there exists three different spatial regions: the majority of the particles 
belong to the first two regions located around one the  two minima and the
remaining particles occupy the third region, the one between the minima.
Intrestingly, the velocity distribution function in the first two regions has a Gaussian form, 
whereas in the third region the velocity distribution
acquires a bimodal shape.
In Section~\ref{section4} we present a theoretical analysis of the model and explain the bimodal behavior
and in Section~\ref{energetics} we briefly discuss the connection between the form of the distribution function and the energetics of the model.
Finally, in Section \ref{section5} we draw the conclusions.


\section{The active Ornstein-Uhlenbeck Model}
\label{section2}

We consider an assembly of non-interacting active particles
in the presence of a confining potential, $U$.  The motion of the particles due to the combined action of the deterministic and active forces is described
by means of the AOUP dynamics, in which the orientations of the particles are 
not explicitly considered and the active force is represented by an Ornstein-Uhlenbeck process.
In two and three dimensions the AOUP is known to capture the same type of phenomenology as the ABP,
which is considered to be a more realistic modeling of active particles. Nevertheless, the AOUP has gained relevance because not only is 
the simplest model sharing with ABP the same two-time autocorrelations and free diffusion 
\cite{farage2015effective} but also because is more convenient for theoretical analysis.

In one dimension,
the AOUP self-propulsion mechanism is assimilated to
a colored noise, $ f^a$ and the governing equations read:
\begin{subequations} \label{eq:AOUPmodel}
\begin{flalign}
\label{eq:AOUPmodel1}
\gamma\dot x&= F(x) + f^a +\gamma \sqrt{2D_t} \xi\\
\label{eq:AOUPmodel2}
\tau\dot f^a& = - f^a + \sqrt{2D_a}\eta
\end{flalign}
\end{subequations}
where $\xi$ and $\eta$ are   white noises,
$\delta$-correlated in time and have
  unit variance and zero mean. It is easy to see that the self-propulsion
force, which is an internal degree of freedom converting energy into
motion, is such that $\langle f^a (t) f^a (t' )\rangle = D_a/\tau
\exp{(-|t-t'|/\tau)} $.
 We fix the ratio $D_a/\tau$, in
order to keep constant the average self-propulsion velocity of one
particle, $\sqrt{\langle ( f^a)^2\rangle} =
\sqrt{D_a/\tau}$.
Hereafter, we consider the limit of strong activity and consequently in Eq.~\eqref{eq:AOUPmodel1} we drop the last term, representing the contribution due to thermal fluctuations.

Experimental studies of  bacterial colonies have shown that $D_a$ can be much larger than $D_t$. For instance, $D_a$ of active bacteria in pure water is about
$100\, \mu m^2/s$, whereas the diffusion coefficient of dead bacteria is approximately $\approx 0.3\, \mu m^2/s$. 
In this and other cases, the contribution due to
the diffusion due to the thermal agitation of the solvent is at least ten times smaller than the one due to the activity
\cite{bechinger2013physics}.
$ F(x) =- U'(x)$ 
is the deterministic force, $U$ the external potential and the prime indicates the spatial derivative. 
As previously shown in Ref.~\cite{caprini2018scirep}, a dimensionless parameter, $\nu$ measures how far from equilibrium is the system: $\nu$  is the ratio between  the persistence time of the trajectory and the relaxation time due to the
external force:  $\nu=\tau
\frac{U''(l)}{\gamma}$
 where $l$ is the typical length of the potential, for instance the effective width of the confining potential. 
 In other words, when $\nu \lesssim 1$, the
relaxation time of the active force is smaller than the typical time
over which a significant change of the microswimmer position, due to
the potential, occurs.
 When
$\nu \ll 1$ the system \eqref{eq:AOUPmodel} 
can be mapped into an overdamped passive system with diffusion $D_a$
and potential $U$, whose behavior is well understood. Indeed, in this
case the activity plays just the role of an effective temperature, or
in other words the distribution is a Maxwell-Boltzmann with
temperature $\gamma D_a$.  Therefore, we restrict our study to the
case $\nu \ge O(1)$, with the aim of studying a far from equilibrium
regime.

In order to make progress, it is useful to map the
Eqs.\eqref{eq:AOUPmodel} onto a Markovian system, transforming from the original
variables $(x,  f^a)$ to the new pair $(x, v)$, where $v=\dot x$. This change
of coordinates \cite{marconi2016velocity} maps the original overdamped dynamics
with colored noise onto the underdamped dynamics of a fictitious passive
particle immersed in a solvent of spatially varying viscosity.  The simultaneous action of deterministic and active 
forces produces a frictional force
$-\gamma v\,\Gamma(x)/\tau $ \cite{marconi2015towards},
which is given by
\be
\Gamma(x) =  1+\frac{\tau}{\gamma} U''(x) \, ,
\label{Biggamma}
\ee
where the double prime symbol stands for second spatial derivative.
 The transformed dynamics
reads:
\begin{eqnarray}
\label{eq:xv_modelx}
&&
\dot{{x}}=v\\
&& \dot{{v}} = -\frac{\Gamma(x)}{\tau} v+ \frac{F(x)}{\tau \gamma}+ \frac{\sqrt{2D_a}}{\tau}\eta .
\label{eq:xv_modelv}
\end{eqnarray}
The statistical properties of the system are described by the probability distribution $p(x,v,t)$ which obeys the following Kramers-Fokker-Planck equation:
\begin{equation}
\begin{aligned}
 \frac{\partial p(x,v,t)}{\partial t}+  v \frac{\partial p(x,v,t)}{\partial x} +\frac{F(x)}{\tau \gamma}\frac{\partial p(x,v,t)}{\partial v}
=   \frac{1}{\tau} \frac{\partial }{\partial v} \left(\frac{D_a}{\tau} \frac{\partial }{\partial v}+\Gamma(x) v   \right) p(x,v,t)
\end{aligned}
\label{fpeq}
\end{equation}

Neither the non-equilibrium dynamics associated with
Eqs. \eqref{eq:xv_modelx} -\eqref{eq:xv_modelv} nor that described by
Eq.~\eqref{fpeq} are easy to solve even in the stationary state. Up to
now, the only known general analytical results for the stationary PDF
have been obtained by an expansion in power of $\sqrt{\tau}$,
i.e. when $\nu \ll 1$ \cite{fodor2016far,marconi2017heat}. In addition, some approximation schemes have
been developed, mainly the so-called UCNA approximation
\cite{hanggi1995colored} (or the Fox approximation
\cite{farage2015effective,wittmann2017effective}). Basically, the UCNA
consists in an adiabatic elimination of the inertial term
\cite{maggi2015multidimensional} in Eq.~\eqref{eq:xv_modelv}, or
equivalently in reducing the Kramers-Fokker-Planck equation
Eq.~\eqref{fpeq} to a Fokker-Planck equation for the positional degrees
of freedom only.  Such a transformation, similar in the spirit to the
reduction from the Kramers to the Smoluchowski representation of the
dynamics \cite{Risken}, leads to an equation which is solved under the
additional assumption of zero currents in the steady state.  It has
been pointed out \cite{fodor2016far}, that in deriving the UCNA
approximation, one invokes the detailed balance (DB) condition, which
states that in equilibrium each elementary process is equilibrated by
its reverse process.  Not surprisingly, the resulting steady state
UCNA probability distribution function, $\rho_{ucna}(x)$ is strikingly
similar to a Maxwell-Boltzmann distribution, with an effective
Hamiltonian which depends on the derivatives of the potential. 
  Notwithstanding this approximation which maps a truly
  non-equilibrium to an equilibrium-like system, in the
  multidimensional and interacting case the UCNA successfully predicts
  some important features of active particles, such as the clustering
  near an obstacle, the tendency of the particles to aggregate, the
  mobility reduction as the density increases. On the other hand,
  there is one caveat which limits the application of UCNA to general
  systems and this is the condition of positivity of the Hessian of
  the interaction potential. When this condition is not fulfilled, as
  we will see later, the UCNA breaks down. First, however, we briefly review the basic facts of the UCNA approximation.

The steady UCNA configurational distribution~\cite{marconi2016effective},  $\rho_{ucna}(x)$, can be interpreted as the marginal distribution of a 
PDF,  $p_a(x,v)$, which approximates the exact time-independent solution of Eq.~\eqref{fpeq}. This approximation can be written as:
\begin{equation}
\label{eq:ER_distribution}
p_{a}(x,v)\approx \rho_{ucna}(x)   \sqrt{\frac{\tau \gamma}{2 \pi \theta(x)}} \exp(-\frac{\tau \gamma v^2}{2\theta(x)} )
\end{equation}
where we introduced the local "temperature"
\be
\theta(x) =
\frac{ D_a \gamma}{1+\frac{\tau}{\gamma}U''(x)} 
\label{kintemperature}
\ee
and
\begin{equation}
\rho_{ucna}(x)\propto \exp(-\frac{H(x)}{D_a\gamma})\, ,
\end{equation}
$H(x)$ being an effective configurational Hamiltonian :
\begin{equation}
H(x)=U(x)+\frac{\tau}{\gamma}U'(x)^2 -\gamma D_a \ln{\left(1+\frac{\tau}{\gamma}U''(x)  \right)} .
\label{effectivepot}
\end{equation}
Hereafter, with the symbol $\langle v^n(x) \rangle$ we shall indicate the conditional average
of the quantity $v^n$ at fixed $x$, i.e.  $\langle v^n|x \rangle $.

Consistently with the UCNA approach the local variance of the velocity, $\langle v^2(x) \rangle $, at a fixed $x$
is approximated by the following formula 
\be
\langle v^2(x) \rangle \approx \frac{\theta(x)}{\tau\gamma} =
\frac{ \frac{D_a}{\tau}}{1+\frac{\tau}{\gamma}U''(x)} \, .
\label{kintemperature}
\ee
Since $\theta(x)$ depends on the
shape of the potential it is position dependent and is constant only for linear and quadratic potentials~\cite{marconi2017heat}.

In principle, when $\tau \gg 1$ there are no a priori reasons to consider
$p_a(x,v)$ as a good approximation, with the exception of quadratic potentials where it is even exact. However, in a recent numerical study \cite{caprini2018scirep} it was shown that for more general potentials the whole configurational space
can be classified in different regions according to the following criterion:
regions where approximation~\eqref{eq:ER_distribution} works, which we name "equilibrium-like regions" (ER), and the remaining "non-equilibrium regions" (NER) where the approximation breaks down. Within the equilibrium-like regions the detailed balance
condition is locally satisfied and the local heat flux approximatively vanishes. In ref.~ \cite{caprini2018scirep} a single-well confining non quadratic potential has been considered: it was found that the peak of the configurational distribution does not necessarily
coincide with the minimum of the confining potential. 
Interestingly, the two symmetric accumulation regions (for a 1D system), far from the potential minimum, are ER.

\section{Phenomenology of an active particle in a double well potential}
\label{section3}
In this Section we study Eqs.\eqref{eq:xv_modelx}-\eqref{eq:xv_modelv} in
the presence of a non-convex potential. Let us consider the following
double-well potential 
\be
U(x)= b\frac{x^4}{4} - a\frac{x^2}{2}
\label{potential}
\ee
which has been intensively studied in the past in the
case of passive brownian particles \cite{mel1991kramers}. 
The escape time of a bistable potential well in a thermal environment is an important problem in physics and chemistry that has been evaluated in various ways. Chemical reactions is a typical example which has motivated many studies.
Kramers, in his seminal paper \cite{kramers1940brownian}, evaluated the stationary escape rate in the high friction regime dominated by a spatial diffusion process.

At equilibrium the basic
phenomenology (at low temperature) is: the particles jump at a random time
from one well to another with a mean time given by Kramers
formula \cite{van1992stochastic}. The same type of problem but with a colored noise replacing the white noise of Kramers treatment was tackled in Ref.~\cite{sharma2017escape}: the authors proposed a generalization of the Kramers formula in the near-equilibrium regime ($\nu \ll 1$) and found that the
escape rate could be derived in terms of an effective potential similar to \eqref{effectivepot}.

In contrast with the assumption of small current adopted in
Kramers' approximation, we consider a regime where the current is
large and the system is far-from-equilibrium.  Therefore, it is clear
from Eq.~\eqref{fpeq} that in addition to the rich phenomenology
determined by the existence of two stable minima, one should also
observe important effects due to the position-dependent effective
friction, $\Gamma(x)$.  In fact, if $U(x)$ is non-convex and $\nu$ is
sufficiently large, $\Gamma(x)$ defined in Eq.~\eqref{Biggamma}
becomes negative whenever $U''(x) < -\gamma/\tau$ and cannot be
considered a friction anymore. In other words, $-\Gamma(x)v/\tau$
instead of acting as a damping force represents an acceleration
forward and in this case $\Gamma^{-1}$ can be interpreted as a
negative mobility.  In the following we explore the physical
consequences of it.

Based upon the
above considerations, we dub ``NTR'' (Negative
Temperatures Region) the zone  where  the effective friction $\Gamma(x)<0$ is negative.
We define $\tau_c$ as the critical value of
$\tau$ such that for $\tau\geq\tau_c$, the friction  $\Gamma(x)\leq0$ is negative for some values of $x$ :\begin{equation}
\tau_c= - \text{min}_x \left( \frac{ \gamma}{U''(x)}\right).
\label{equotto}
\end{equation}
In the case of the potential~\eqref{potential} 
 the critical value is $\tau_c = \frac{\gamma}{b}$ and
the system exhibits a sort of bifurcation depending on
$\tau$: if $\tau<\tau_c$  the temperature $\theta(x)$  of the system is positive everywhere, while for
$\tau\geq\tau_c$ the temperature becomes negative in some region located around the maximum of the potential.
We determine the size of the interval, $(-x_N, x_N)$, where $\theta(x)\leq0$
 by considering the solutions of Eq.~\eqref{equotto} with $\tau_c$ replaced by a fixed value of $\tau$:
$$|x_N| = U''^{-1}(-\frac{\gamma}{\tau}) ,$$ 
where $U''^{-1}$ is the inverse
function of the second derivative of $U(x)$. Intuitively, increasing $\tau$ corresponds to enlarge the NTR, until 
a saturation length is reached.  In particular, for our potential
choice we obtain:
\begin{equation}
\lim_{\tau \to \infty} |x_N| = \lim_{\tau\rightarrow \infty}\sqrt{\frac{b}{3a}(1-\frac{\gamma}{\tau b} )}=  \sqrt{\frac{b}{3a}}
\end{equation}
Thus  for large $\tau$, $x_N$ coincides with the binodal line (the locus of $U''(x)=0$), associated with the potential $U(x)$.

It is interesting to mention that in equilibrium statistical mechanics one may find absolute negative temperatures in the study of some Hamiltonian models, for instance a system of heavy rotators immersed in a bath of light rotators  \cite{cerino2015consistent,puglisi2017temperature}.
In these Hamiltonian systems the occurrence of an absolute negative temperature is a consequence of the form of the kinetic part of the Hamiltonian which is not the standard quadratic function of momenta but rather a periodic function of them. This unusual fact does not rule out the possibility of formulating in a consistent way a Langevin equation for the slow variables of the problem, in this case the momenta of the heavy particles. The occurrence of a negative temperature in the Langevin effective equation is related to the sign change of the Stokes Force~\cite{baldovin2018langevin}.

\begin{figure}[!t]
\centering
\includegraphics[width=0.85\linewidth,keepaspectratio]{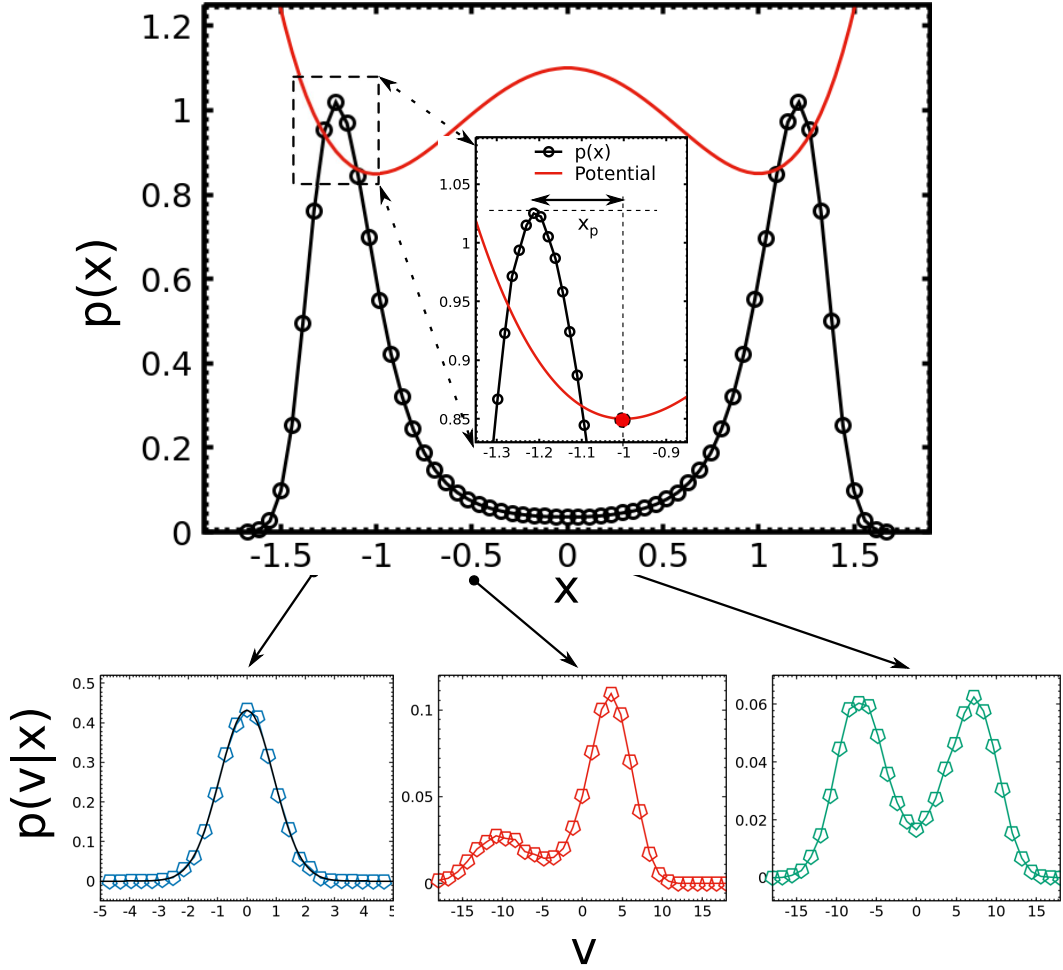}
\caption{Upper panel: The marginal distribution $\rho(x)$ (black data) obtained from the numerical data and the double well potential (the red curve, rescaled for presentation reasons) for the parameter choice $a=b=10$ and. In the inset, we
zoom the distribution near its left maximum. 
 The variance of the self-propulsion force is $D_a/\tau=10^2$ and $\tau=10$. Notice  that the positions of the peaks  of the density $\rho(x)$  are slightly shifted with respect to the minima of the double well potential \eqref{potential} and correspond fairly well to the roots of \eqref{thetaequa}. 
    Lower panel: we display the conditional 
  velocity distribution, $p(v|x)$,
  for three different positions, $x=-1.2, -0.5, 0.$ from the left to the
  right, respectively.
 In the vicinity of the two peaks of $\rho(x)$ at $x=\pm 1.2$ the distribution is unimodal $p(v|x)$.
 At $x=\pm 0.5$ a lateral shoulder appears as a clear indication of the onset of bimodality in the velocity distribution.
 Finally, at the symmetric point $x=0$ the bimodality of $p(v|x)$ is fully developed and the particles form two distinguishable populations: one of particles propagating towards the right and the other towards the left. 
  }
  \label{fig:pdf}
\end{figure}

Hereafter, we present some results obtained by solving numerically the
stochastic differential equations \eqref{eq:xv_modelx}
-\eqref{eq:xv_modelv} the Euler-Maruyama algorithm
\cite{toral2014stochastic}.  Let us start showing in the top panel
Fig.~\ref{fig:pdf} the marginal space density, $\rho(x)$.  One may
observe that the shape of the spatial distribution is qualitatively
similar to the one we would find in the case of passive Brownian
particles: two high density regions appear near the side minima.
Considering the profile $\rho(x)$, the most relevant difference with
respect to a Brownian system is represented by the shift of the peaks
with respect to the location of the potential minima.  The shift can
be estimated by imposing the balance between the deterministic force,
$F(x)$, and the active force, $f^a$, which we roughly approximate as
$\sim \pm \gamma \sqrt{D_a/\tau}$, taking the plus sign for particles
belonging to the left minimum and the minus to those belonging to the
right one. This approximate treatment of $f_a$ works
better in the regime of large $\nu$, whereas in the small $\tau$
regime the active force has to be considered as a white noise, as
discussed in section \ref{section2} and therefore does not contribute to
the force balance.  For the potential \eqref{potential}, the two peaks
are given by the real solutions of the following equation:
\begin{equation}
-ax^3+b x =  \gamma\sqrt{\frac{D_a}{\tau}} [\Theta(-x)-\Theta(x)]
\label{thetaequa}
\end{equation}
where $\Theta(x)$ is the Heaviside step function.
For $x>0$  we take the largest root, while
for $x<0$ we take the most negative root and these roughly correspond to the maxima of the PDF.
One can see that the distance between the positions of the two maxima increases with the ratio $D_a/\tau$,
in agreement with Eq.~\eqref{thetaequa}.

Now, we focus our analysis on the very active regime where $D_a/\tau$
is large enough so that we have frequent jumps between wells.
On the contrary, when $D_a/\tau$ is too small the jumps are rare.  A
criterion to determine this threshold value is to impose the condition
that the mean amplitude of the active force, $\gamma\sqrt{D_a/\tau}$,
exceeds the maximal value of the repulsive force $F(x)$ in a region
contained between the minimum and the central maximum of the
potential.


\begin{figure}[!b]
\centering
\includegraphics[width=1\linewidth,keepaspectratio]{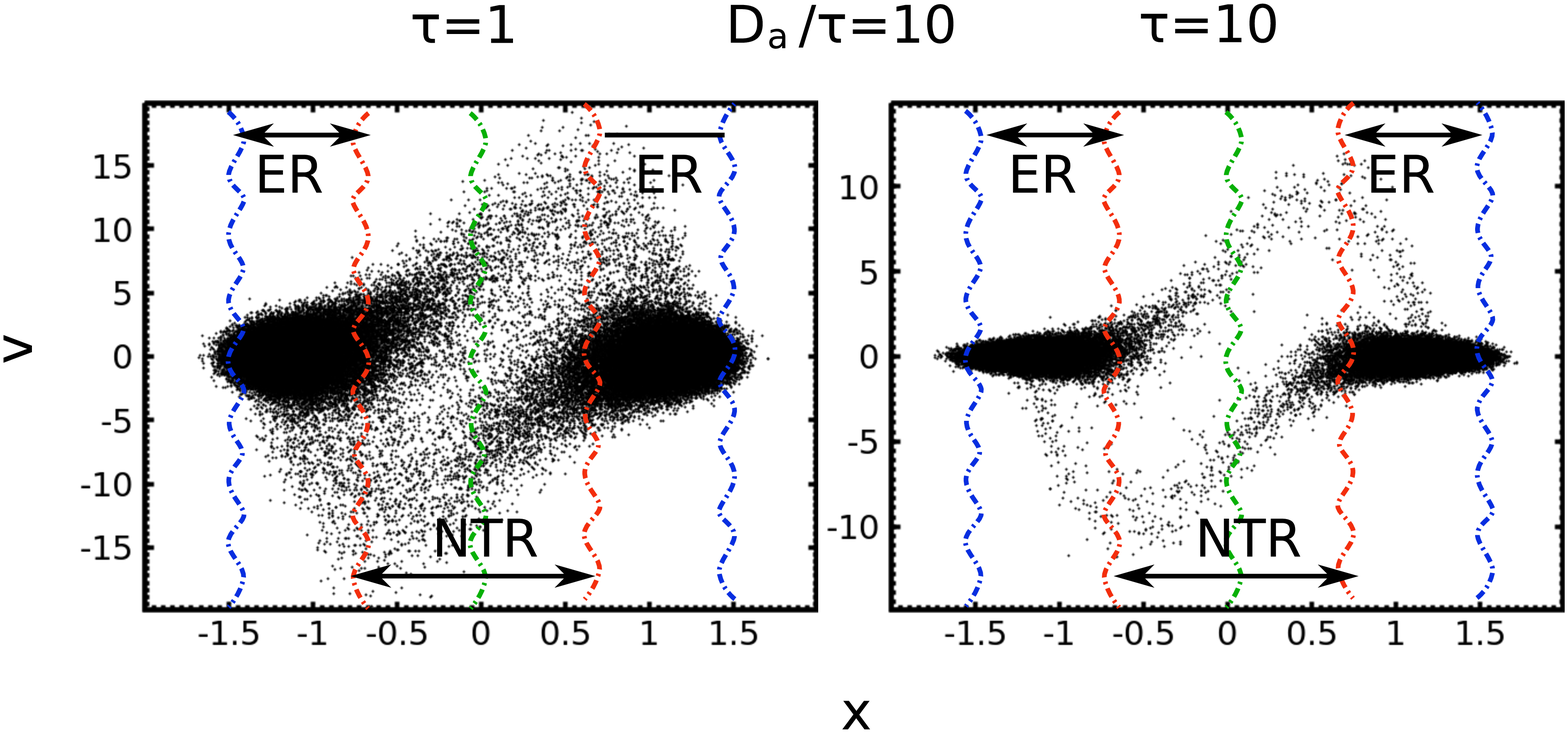}
\caption{ We show the snapshots representing the phase-points in the $(x,v)$-plane
for two different choices of the persistence time $\tau$: left panel $\tau=1$ and right panel  $\tau=10$. The remaining parameters are $D_a/\tau=10^2$ and $a=b=10$. Each point represents the state of one system at a given instant.
We simulated a collection of $N=10^3$ particles. The vertical dashed curves are only a guide to the eye.  
}
\label{fig:phasespace}
\end{figure}


As we said above,
 the overall structure of $\rho(x)$ is not too dissimilar from the one one would observe in a thermal system,
but if we scrutinize the system using a different indicator, i.e. the
conditional probability of the system's ``velocity'', $p(v|x) =p(x,v)/\rho(x)$, the peculiarity of the active system becomes evident.  
In  the bottom panels of Fig.~\ref{fig:pdf} we plot the numerical data representing the
conditional probability of the system's ``velocity'', $p(v|x)$ for three different typical positions,  $x$:
at the $\rho$-peak position, at the position of the maximum of the potential and at the intermediate point between them (left,right and middle panel, respectively).
In the bottom left panel of Fig.~\ref{fig:pdf}, the conditional velocity distribution displays a single peak: such a form is consistent with the distribution Eq.~\eqref{eq:ER_distribution} and corresponds
 with the existence of a positive local kinetic temperature which is well reproduced by
  $\theta(x)$ given by \eqref{kintemperature}.
 We classify such a behavior as equilibrium-like and dub these regions of phase-space as ER. 
  We define as "non-equilibrium regions" (NER) the zones which are not ER. In particular, the NTR defined above is contained in the NER, since the study of the $p(v|x)$ shows clear non-equilibrium features. Indeed, for those values of $x$ where $\Gamma(x)$ becomes negative, the observed scenario is now more intriguing because the conditional distribution $p(v|x)$ displays bimodality: in the bottom center panel
   for $x=-0.5$ the main peak is shifted from $v=0$ to a positive 
  value $v\approx 5$ and is accompanied by the emergence of a second lower peak centered at
  negative $v\approx-10$ (and vice-versa on the other side).  Such an unbalanced shape of the $p(v|x)$ distribution disappears for $x=0$ where the two peaks are symmetric ( see bottom right panel). The $p(v|x)$ shown in
 the central and right panels is not consistent with Eq.~\eqref{eq:ER_distribution} and cannot be accounted for
 by a Maxwell-type distribution.

In the two panels of Fig.~\ref{fig:phasespace}, we display two phase-space
snapshot configurations in the plane $(x,v)$ 
 for two different choices of $\tau$ ($\tau=1,10$).
In order to sample configurations corresponding to the same strength of the active force
we used the same ratio
$D_a/\tau=10^2$.  Both snapshots reveal an
interesting phenomenology which confirms the previous
 observations and helps us to gain some insight: particles spend most of their life in the equilibrium-like regions, but sometimes visit the NTR region. 
  When this occurs, the particles experience an acceleration, due to the negative mobility $1/\Gamma(x) <0$, towards the opposite ER region.
 Finally, when the particles have crossed the NTR and reached the opposite side $\Gamma(x)$ returns to positive values and the motion becomes damped again with a concomitant restoring of the local Gaussian velocity distribution. 
 
 The $(x,v)$ scatter plot reveals the presence of two lanes (one in the in the upper half-plane and the other 
  in the lower half)
 of representative points  in the NTR connecting the two darker regions.
 If $\tau$ increases the two lanes become thinner along the $v$ direction as clearly shown in the right panel of
 Fig.~\ref{fig:phasespace}.
 Let us also remark that a larger value of the persistence time, $\tau$, determines a stronger
 selection mechanism of the velocities of the particles which succesfully escape from one well to the other.

\begin{figure}[!t]
\centering
\includegraphics[width=1\linewidth,keepaspectratio]{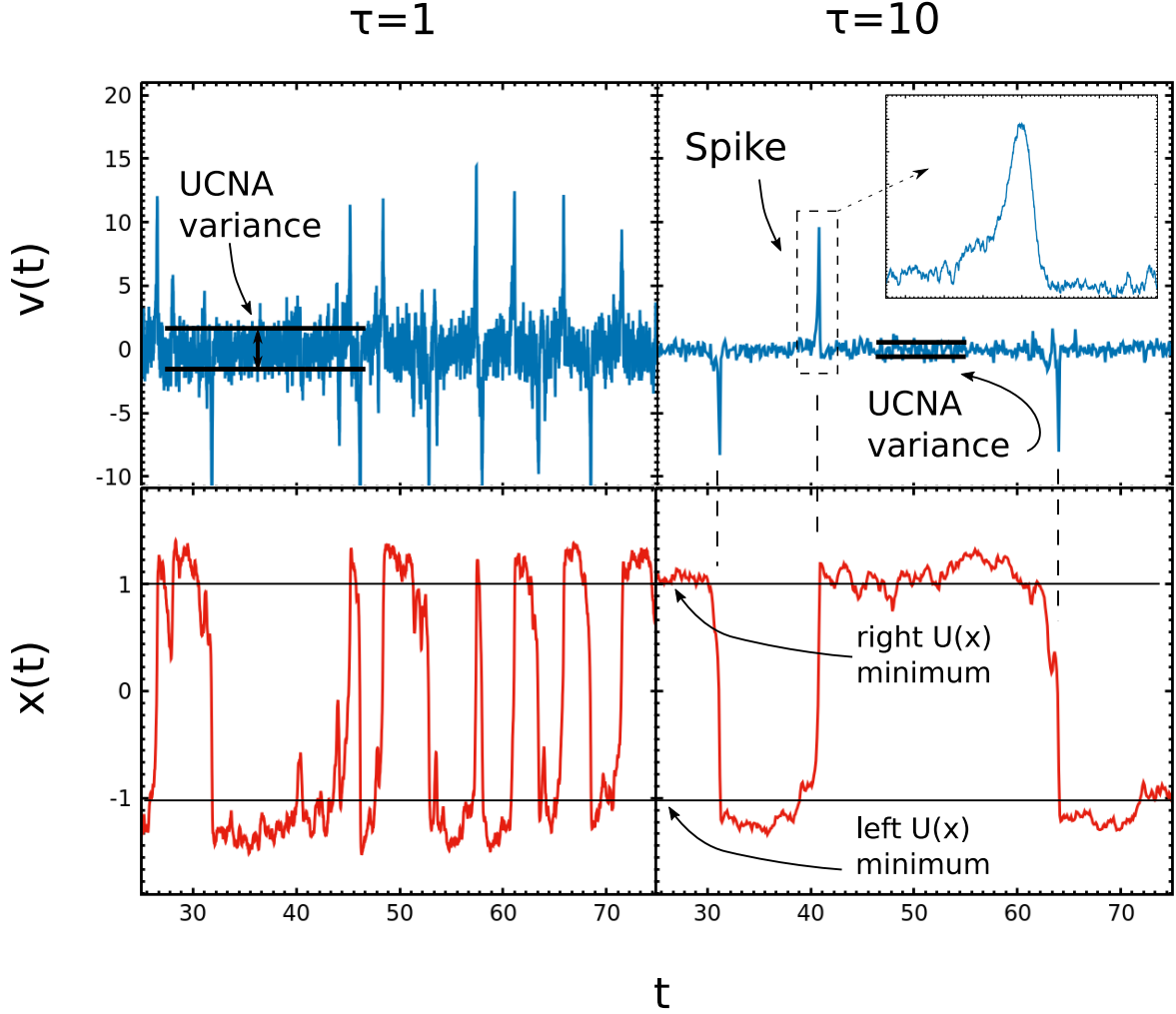}
\caption{In the upper left and right panel we show the evolution of the 
position, $x(t)$ for $\tau=1$ and $\tau=10$, respectively. 
In the lower panels, we display the corresponding velocities. 
The values of the remaining parameters are: $D_a/\tau=10^2$ and $a=b=10$.}\label{fig:xvt}
\end{figure}

 Such a picture is also confirmed by Fig.~\ref{fig:xvt} where we report the values $v(t)$ and
  $x(t)$ for a single particle trajectory: in correspondence of the instant when the particle changes well, its velocity rapidly grows, a
  scenario which has not a Brownian analog and is peculiar of the active dynamics.
  In particular, $v(t)$ reveals pronounced spikes at the crossing barrier time, whose height are consistently larger that  the typical equilibrium-like fluctuation, predicted by UCNA. The shape of the spikes resembles a deterministic trajectory and thus does not seem particularly affected by the random force, as shown in the Inset of Fig.~\ref{fig:xvt}.

Summarizing, we have found regions characterized by
``local'' negative temperatures: when a particle enters one of these regions, it becomes too energetic to remain there and thus has to leave it to enter again
an ER, where $\Gamma(x)$
becomes again positive. This feed-back mechanism  explains why the mean velocity 
does not grow indefinitely.

\section{Analysis of the Negative Temperature Region}
\label{section4}
As shown in Figs.~\ref{fig:pdf} and \ref{fig:phasespace}, an active particle spends the most 
of its life in the regions where the potential $U(  x)$ is convex.  
 Under such condition, the UCNA
approach is expected to work and this fact is confirmed by the observation that the numerical
$p(v|x)$ is well reproduced by a Gaussian with temperature
$\theta(x)$ given by Eq.~\eqref{kintemperature}. However, this approximation is not uniformly valid in space
as the presence of two peaks in $p(v|x)$ of Fig.~\ref{fig:pdf} and the form of the snapshots of Fig.~\ref{fig:phasespace} have shown. 
To remedy this situation,
in the present Section we attempt to formulate in alternative to the UCNA
a theoretical explanation  of the observed behavior in the region $(-x_N,x_N)$
i.e.  where $\Gamma(x)<0$.

Hereafter, we propose a theoretical interpretation of the numerical findings illustrated above and is based on the analysis of the behavior of the slow variables  of the system in this unstable region. 
We choose as slow variables the average velocity of the particles and velocity variance at fixed $x$ ( i.e. conditional averages) and determine their evolution.
Since
in the stationarity state the average velocity at a fixed arbitrary position vanishes, i.e.
$\langle v(x)\rangle={\int dv  v p(x,v) }/{\int dv p(x,v)}=0 $,
it is crucial to study separately the population of the particles going from the left well to the right one
(the upper lane in Fig.~\ref{fig:phasespace}),
from those performing the opposite path.

 To identify these two populations we define $p_{+}(x,v)$ ($p_{-}(x,v)$) as the unnormalized PDF of the particles whose dynamics starts
from the left (right) well. In practice, we compute $p_{\pm}$ counting the particles in the two lanes of Fig.~\ref{fig:phasespace} and normalizing them with the total number of particles. In this way, summing over the two populations we recover the total distribution:
$$
p(x,v)=p_{+}(x,v)+p_{-}(x,v).
$$
The functional form of $p_{\pm}$ in the NTR will be investigated in the present section. Accordingly, we define the first  conditional velocity moment with respect to $p_{\pm}(x,v) $:
$$\langle v(x) \rangle_\pm=\frac{\int p_\pm(x,v) v dv }{\int p_\pm(x,v) dv }.$$  


 Now,  averaging Eq.~\eqref{eq:xv_modelv}  with respect to $p_{\pm}(x,v) /\int dv p_{\pm}(x,v)$ we obtain:
\begin{flalign}
\label{eq:av_velocity}
\frac{d}{dt}\langle v(x) \rangle_\pm = - \frac{1}{\tau}\left(1+\frac{\tau}{\gamma} U''(x)\right)\langle v(x) \rangle_\pm - \frac{U'(x)}{\gamma \tau} .
\end{flalign}
In the region, where $\Gamma(x)<0$, the absolute value of the average $\langle v(x)\rangle_\pm$ grows exponentially in time before 
dropping to zero again, as one can observe qualitatively the inset of top-right panel of Fig.~\ref{fig:xvt}. In order to 
determine $\langle v(x) \rangle_\pm$, we specialize the treatment to the regime $\gamma\tau \gg 1$,  where
some approximations are possible and
we can neglect the  last term in the right hand side of the Eq.~\eqref{eq:av_velocity}. 
This regime is the most interesting for the present study, since it is just the activity dominated regime. 
The integration of equation \eqref{eq:av_velocity} is performed taking into account that the conditional averages $\langle v(x)\rangle_\pm$ are explicit functions of $x$ and $t$ so that the operator $d/dt$ is the total derivative:
$$d/dt  = \partial_t + v\partial_x \approx \partial_t + \langle v(x) \rangle_\pm \partial_x  \, .$$
 We integrate  Eq.~\eqref{eq:av_velocity} and obtain:
\begin{equation}
\label{eq:average_prediction}
\langle v(x) \rangle_\pm = \langle v (x_0)\rangle_\pm - \frac{U'(x)}{\gamma} + \frac{U'(x_0)}{\gamma} + O\left(\frac{1}{\tau} \right)
\end{equation}
where the lower limit of integration $x_0$ is given by  $ -x_N$ if the initial configuration 
starts in the left sector  and the particle propagates from the left towards the right, while we choose $x_0=x_N$ in the opposite case.
Such an equation expresses the conservation of the average self-propulsion, since $\langle f_a(x_\pm ) \rangle=\gamma \langle v(x) \rangle_\pm + U'(x) =\pm const$ for $x \in (-x_N, x_N)$.
Since we are considering $\tau\gg1$ the microswimmers able to reach the NTR, are the fraction of particles with self-propulsion, $f_a$, large enough to reach the point $x_0$, as discussed in Section III. Moreover, until this fraction of particles reaches the opposite well,
their self-propulsion will maintain a nearly constant value.

Let us, now, consider the equation of evolution for the slow variable $\langle v^2(x) \rangle_\pm $.
To this purpose, we multiply by $v$  Eq.~\eqref{eq:xv_modelv} and  apply the Ito calculus \cite{gardiner2009stochastic} and obtain:
\begin{equation}
\begin{aligned}
vdv =  -\frac{\Gamma(x)}{\tau} v^2 dt + v\frac{\sqrt{2D_a}}{\tau} \,dw  - v\frac{U'(x)}{\tau\gamma}  dt   
\end{aligned}
\end{equation}
where $dw$ is the Wiener process associated with the white noise, $\eta$.
By taking the average, $\langle \cdot\rangle_\pm$, we write an equation for $\langle v^2 (x)\rangle_\pm$, which depends on $\langle v(x) \rangle_\pm$:
\begin{equation}
\label{eq:v2_eq}
\frac{d}{dt}\langle v^2(x)\rangle_\pm=  \langle v(x) \rangle_\pm\frac{d\langle v^2(x)\rangle_\pm}{dx}=-2\frac{\Gamma(x)}{\tau} \langle v^2(x) \rangle_\pm - 2\langle v(x)  \rangle_\pm \frac{U'(x)}{\tau\gamma}  +  2\frac{D_a}{\tau^2}
\end{equation}
where again in the first equality $d/dt$ is the total derivative.
Such an equation, in the case of vanishing currents ($\langle v(x)\rangle_\pm=0$) and positive temperature, is in agreement with the UCNA prediction because we have the simple result:
$
\langle v^2(x) \rangle_\pm=\frac{D_a}{\tau \Gamma(x)}, 
$
which is nothing but formula~\eqref{kintemperature}.
In the more interesting case of non vanishing currents, Eq.~\eqref{eq:v2_eq} is an inhomogeneous first order differential
equation for the observables $\langle v^2(x) \rangle_\pm$ and can be integrated through the formula
\begin{equation}
\begin{aligned}
\langle v^2(x) \rangle_\pm &=  \langle v^2(x_0) \rangle_\pm e^{-\int^x_{x_0} dx' g(x')}  \\
&+ e^{-\int^x_{x_0} dx' g(x')} \int^{x}_{x_0} dx' T(x') e^{\int^{x'}_{x_0} dx'' g(x'')}
\end{aligned}
\end{equation}
where $g(x)= 2\frac{\Gamma(x)}{\tau \langle v(x)\rangle_\pm}$, $T(x)=\frac{2 D_a}{\tau^2\langle v(x) \rangle_\pm}$, 
and the term $\frac{U'(x)}{\tau\gamma}=O(1/\tau)$ and has been neglected. 
In the limit $\tau \gg 1$, we obtain:
\be
\int dx  g(x)=
\int dx \frac{2 \Gamma(x)}{\left(\tau \langle v(x) \rangle_\pm\right)} \approx -2 \ln{\langle v(x)\rangle_\pm }+ O(1/\tau)
\ee
so that the solution of the inhomogeneous first order ordinary differential equation\eqref{eq:v2_eq}, simply reads:
\begin{equation}
\label{eq:v2_eq3}
\langle v^2(x)\rangle_\pm \approx \langle v^2(x_0) \rangle_\pm \left(\frac{\langle v(x)\rangle_\pm}{\langle v(x_0)\rangle_\pm}\right)^2 + \left(\langle v(x)\rangle_\pm\right)^2\int^x_{x_0} dx' \frac{ 2 D_a}{\tau^2 \left(\langle v(x')\rangle_\pm\right)^3 }.
\end{equation}
Finally, since the last term in the the right hand side of Eq.~\eqref{eq:v2_eq3} is negligible, being $O(D_a/\tau^2)$  the variances $\Delta_\pm(x)\equiv \langle(v-\langle v(x)\rangle_\pm)^2 \rangle_\pm$ can be approximated as:
\begin{equation}
\label{eq:variance_prediction}
\Delta_\pm(x) \approx \left(\langle v(x) \rangle_\pm\right)^2 \frac{\Delta_\pm(x_0)}{ \left(\langle v(x_0) \rangle_\pm\right)^2}.
\end{equation}

Eq.~\eqref{eq:variance_prediction} establishes a simple approximate relation between the variances, $\Delta_\pm(x)$, and the first conditional velocity moments, $\langle v(x) \rangle_\pm$, which holds in the regime $\tau \gg 1$. 
We can roughly estimate 
the velocity variance at $x_0$ as $\Delta_\pm(x_0) \approx D_a/(\tau \Gamma(x_0))\sim D_a/(\tau^2 U''(x_0))$
and explain why at fixed $D_a/\tau$ the lanes in the scatter plot of Fig.~\ref{fig:phasespace} become thinner as $\tau$ increases.

\begin{figure}[!t]
\centering
\includegraphics[width=1\linewidth,keepaspectratio]{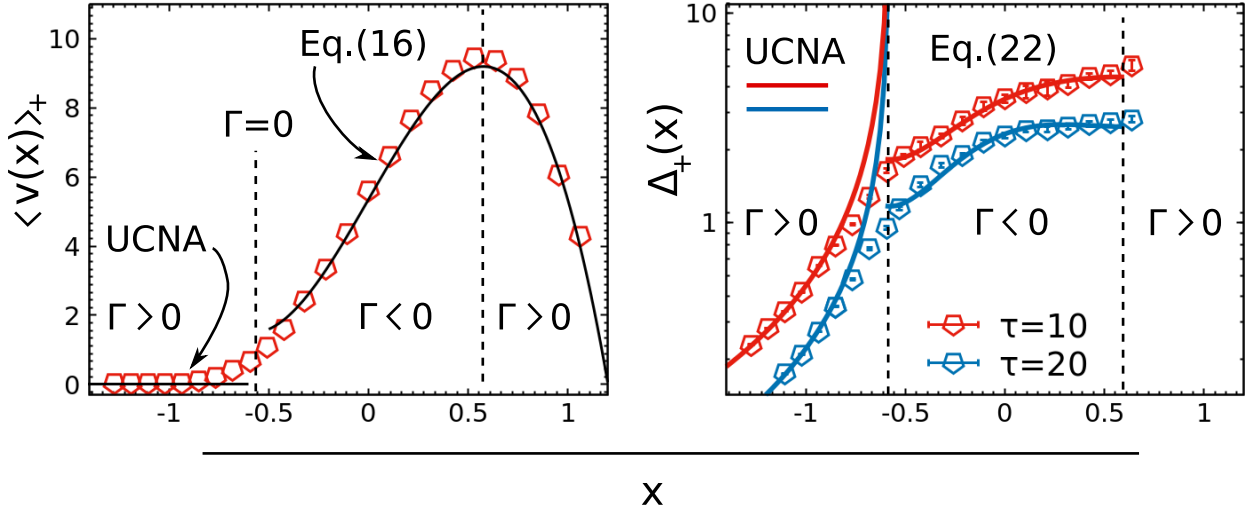}
\caption{Left Panel: $\langle v(x) \rangle_{+}$ computed from data (red points) and from the theoretical prediction (black line) of Eq.~\eqref{eq:average_prediction} with the initial value, $\langle v(x_0) \rangle_{+}=1.5$, obtained from the numerical data values data in correspondence of $\tau=10$. Right Panel: Variance $\Delta_{+}(x)$ computed from data (points) and from the theoretical prediction, respectively:  in the left sector  ($\Gamma>0$) the continuous lines represent the
UCNA theoretical value for $\tau=10$ and $20$.
In the central region ($\Gamma<0$) the continuous lines represent the prediction of  Eq.~\eqref{eq:variance_prediction}  .  The remaining parameters are $D_a/\tau=10^2$, $a=b=10$.}\label{fig:Moments}
\end{figure}

Let us consider now  the phase-space distribution, $p(x,v)$ in the central region: using the above results we see that
 it can be approximated as the sum of two Gaussians, representing, respectively, the left and the right population:
\begin{equation}
\label{eq:NER_pdf}
p(x,v) \approx\mathcal{N} \left( \rho_{+}(x)   \sqrt{\frac{1}{2 \pi \Delta_+(x)}}  \exp\left(-\frac{\left(v-\langle v(x) \rangle_{+} \right)^2}{2 \Delta_{+}(x)}\right) + \rho_{-}(x) \sqrt{\frac{1}{2 \pi \Delta_-(x)}}\exp \left(-\frac{\left(v-\langle v(x) \rangle_{-} \right)^2}{2 \Delta_{-}(x)} \right) \right)
\end{equation}
being $\mathcal{N}$ the normalization of the whole distribution. The form of \eqref{eq:NER_pdf} reproduces the
structures observed in the lower panels of Fig.~\ref{fig:pdf} and clearly, the presence of the double peak in the velocity distribution is the signature of a departure from the global equilibrium condition.

The theoretical predictions for $\langle v (x)\rangle_\pm$  and  $\langle v^2 (x)\rangle_\pm$ are tested against
the respective numerical measures
as shown in the left and right  panel of Fig.~\ref{fig:Moments}, respectively.
In particular, the study of $\langle v(x) \rangle_{+}$  obtained from data with the prediction \eqref{eq:average_prediction} displays a good agreement in the NTR region, which confirms the present theory. Notice that Eq.~\eqref{eq:average_prediction} does not hold in the region where $\Gamma(x)>0$ and $\langle v(x)\rangle_\pm \approx0$,  which is instead well described by the UCNA approximation.

 The variance of the left population, $\Delta(x)_{+}=\langle v^2(x)\rangle_{+}-\left[\langle v(x)\rangle_{+}\right]^2$,
shown in the right panel of Fig.~\eqref{fig:Moments},  increases monotonically  with $x$. 
In particular,  the variance reveals a good agreement with the UCNA prediction in the regions where $\Gamma(x)>0$. 
As  $x$ tends from the left  to  $-x_N$, the space point where $\Gamma(x)=0$, the variance calculated with the UCNA diverges, in clear disagreement with the numerical findings of Fig.~\ref{fig:Moments}.
 However, if we consider the NTR region the agreement  between data and the theoretical Eq.~\eqref{eq:variance_prediction} is pretty good, confirming the validity of the approach.


\begin{figure}[!t]
\centering
\includegraphics[width=1\linewidth,keepaspectratio]{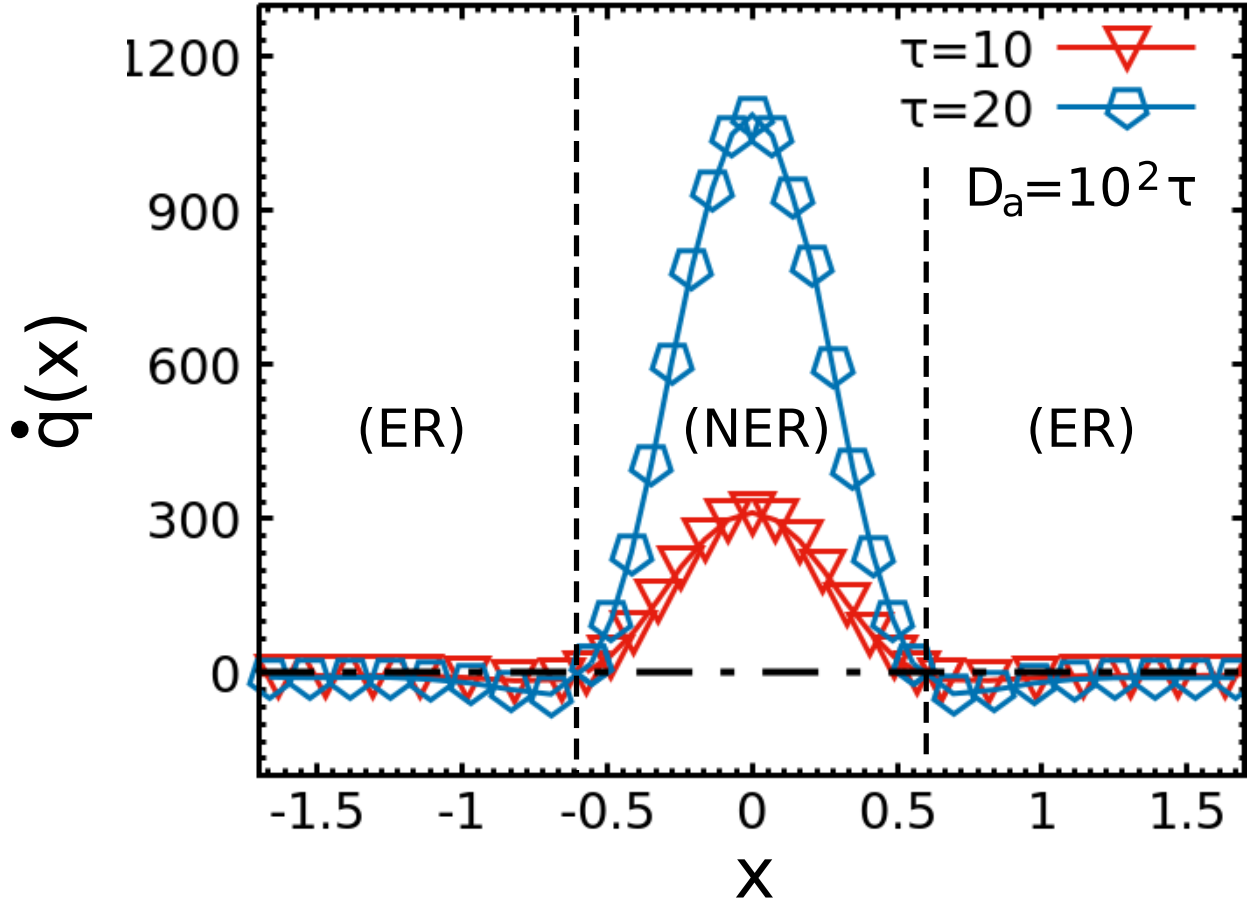}
\caption{
Plot of the heat flux, $\dot{q}$, defined by Eq.~ \eqref{tildeq}  as a function of the position for two different values of $\tau=10,20$ and $D_a/\tau=10^2$. Notice that $\dot{q}$ is quite small and negative in the equilibrium-like region and large and positive in the negative temperature region. The main contribution to the entropy production stems from the equilibrium like region because the majority of the particles sit there. The remaining parameters are  $a=b= 10$.
}
 \label{fig:heatrate}
\end{figure}


\section{Energetics}
\label{energetics}
We conclude with a brief discussion of the energetics of the
model.  Following the methods of Ref. \cite{marconi2017heat} we define the local heat flux,
$\dot{q}(x,t)$, as  the energy flux transferred from the active
 bath  (represented by the Gaussian colored
noise) to the particles. 
There it was shown, that such a flux can be expressed as:
\begin{equation} \label{tildeq}
\dot q(x,t) =- \frac{\Gamma(x)}{\tau \rho(x,t)}\int dv \Bigl[\tau \gamma v^2 p(x,v,t) + \theta(x) v\frac{\partial}{\partial v} p(x,v,t) \Bigr ]  =  -\frac{1}{\tau}  \frac{D_a \gamma}{\theta(x)}  \left[\tau\gamma\langle v^2(x)\rangle -\theta(x)\right ].
\end{equation}
and that the entropy production towards the surrounding  medium reads:
\be
\dot s_m(x,t)=  \frac{\dot q(x,t)}{ \theta(x) }  .
\label{localentropy}
\ee 
It was also demonstrated that the total entropy production  of the medium $\dot S_m(t)=\int dx \rho(x,t) \dot s_m(x,t)$ and the total heat flux are related through a generalised Clausius inequality~\cite{marconi2017heat}
\begin{equation}
\dot S_m(t)=\int dx \rho(x,t)\frac{1}{\theta(x)} \dot{q}(x,t) \le 0.
\label{clausius}
\end{equation}
Such an expression for the entropy production of the AOUP is consistent with the results of Refs.~\cite{fodor2016far,puglisi2017clausius,dabelow2018irreversibility,caprini2018comment}.

 Let us discuss the behavior of $\dot{q}(x, t)$ in the steady state, as shown in Fig.~\ref{fig:heatrate}. 
We can identify two symmetric space regions, occurring at $\Gamma(x)>0$, where $\dot{q}(x)$ (or equivalently $\dot{s}_m(x)$) is almost zero. These zones coincide with the "equilibrium-like regions" (ER) defined in Sec.~\ref{section3}, a nomenclature which is well justified also from a stochastic thermodynamic approach. To be precise, in this region $\dot{q}(x)$ is not exactly zero but assumes very small negative values because in the last equality of Eq.~\eqref{tildeq} we have $\langle v^2(x)\rangle  \geq \theta(x)/\tau \gamma$ and thus the contribution to $\dot{S}_m$ is negative. On the other hands, we call "non-equilibrium space region" (NER) those zones where $\dot{q}(x)$ is large. In particular, the NTR region, introduced in Sec.~\ref{section3}, is strictly contained in the NER, displaying large $\dot{q}(x)$, which assumes positive values. Indeed, both terms $-\theta(x)$ and $\langle v^2(x)\rangle$ contained in the last equality in Eq.~\eqref{tildeq} have the same positive sign.
As shown in Fig.~\ref{fig:heatrate}, $\dot{q}(x)$ grows as $\tau$ increases. Indeed, evaluating  the regime $\tau\gg1$ in Eq.~\eqref{tildeq} in the NTR, $\dot{q}(x)\approx -\Gamma(x)\langle v(x) \rangle \sim \tau |U''(x)| \langle v^2(x)\rangle$, showing the occurrence of an explicit $\tau$ dependence.
Despite $\theta(x)$ approach to infinity for $\Gamma(x)=0$, in this special case $\dot{q}(x)$ is finite and positive and $\dot{s}_m(x)$ becomes zero. 
We outline that the correct sign of the inequality \eqref{clausius} is
realized because $\rho(x)$ is very small in the NER and large in the ER.

Finally, we discuss the connection between the form of the distribution functions and the detailed balance condition.
We have mentioned that in order to derive the UCNA steady state distribution one has to assume the vanishing of
all currents, which is a form to say that the detailed balance condition holds. However, considering the AOUP
one sees that the detailed balance condition holds and the entropy production vanishes only in the case of linear or
quadratic potentials. For more general potentials the DB does not hold. Now, we argue that in the ER the local entropy production \eqref{localentropy} is nearly vanishing, while in the NER (and in particular in the NTR) the local entropy production is large. In the first case a local version of the detailed balance appears to be satisfied whereas in the second case it is strongly violated. This picture is consistent with the Fig.~\ref{fig:pdf}: the Gaussian form of the $p(v|v)$ distribution shown
in panel (b) of Fig.~\ref{fig:pdf} indicates that the detailed balance condition is satisfied locally in the ER, whereas in the NTR
(panels (c) and (d)) the presence of two peaks clearly shows a breakdown of such a condition. 

\section{Conclusion}
\label{section5}
Some comments are in order: 
we have studied an active Ornstein-Uhlenbeck particle in the presence of a bistable potential and found that
for sufficiently large values of the persistence time $\tau$ 
 the space accessible to the particle can classified into regions where the sign of the friction function $\Gamma(x)$
is either positive or negative, that we named equilibrium-like regions (ER) and nonequilibrium regions (NER), respectively.
In the ER, characterized by a small entropy production and by the absence of currents,
 the statistical properties of the system are captured fairly well by an extended UCNA approximation
 which predicts a steady state unimodal distribution function $p(x,v)$ of the Maxwell-Boltzmann type.
 On the contrary, the NER is characterized by the bimodality of the velocity distribution,
 by larger values of the entropy production and by a strong departure from the detailed balance condition.
 Our theory, which is valid in the limit of large $\tau$, succesfully explains the dependence of the local currents $\langle v(x)\rangle_\pm$ and velocity moment $\langle v^2(x)\rangle_\pm$ on the control parameters.
 
We envisage an interesting application of our study: by employing a non-convex potential we may find a velocity selection mechanism which allows us to produce particles with a particular velocity. The selection becomes more efficient as the  persistence time at fixed propulsion speed increases. Moreover, this "device" gives the possibility of producing active particles with a super speed, $|v_s|$, some order of magnitude larger than the typical velocity of particles in the potential-free region.  Clearly, this effect could be amplified by choosing the potential in such a way that $\Gamma(x)$ becomes more negative.
It is worth to mention the fact that the presence of negative mobility regions~\cite{eichhorn2002brownian} is even more severe in two and three dimensions
where it naturally occurs for instance near a concave surface~\cite{fily2017equilibrium}. 
In this case, the mobility is a tensor and its tangential
components may become negative when the curvature radius is small. 
Another important issue is the case where the particles are mutually interacting via some pair potential: the mobility matrix even for small persistence can display negative eigenvalues in such a way that the UCNA is not applicable
in all regions.
It would be interesting
to explore if these multidimensional cases could be treated using concepts similar to those exploited
in the present work.

\acknowledgments
The authors acknowledge fruitful discussions with Marco Baldovin.



%

\end{document}